# Spot size measurement of flash-radiography source utilizing the pinhole imaging method

WANG Yi (王毅), LI Qin(李勤), CHEN Nan(陈楠), CHENG Jinming(程晋明), XIE Yutong(谢宇彤), LIU Yulong(刘云龙), LONG Quanhong(龙全红)

Key Laboratory of Pulsed Power, Institute of Fluid Physics, China Academy of Engineering Physics, Mianyang 621900, Sichuan Province, China

**Abstract:** The spot size of the x-ray source is a key parameter of a flash-radiography facility, which is usually quoted as an evaluation of the resolving power. The pinhole imaging technique is applied to measure the spot size of the Dragon-I linear induction accelerator, by which a two-dimensional spatial distribution of the source spot is obtained. Experimental measurements are performed to measure the spot image when the transportation and focusing of the electron beam is tuned by adjusting the currents of solenoids in the downstream section. The spot size of full-width at half maximum and that defined from the spatial frequency at half peak value of the modulation transfer function are calculated and discussed.
**Key words:** spot size, x-ray source, pinhole, linear induction accelerator

## 1. Introduction

Flash radiography plays a significantly important role in the research of hydrodynamic experiments, which enables recording of inner images for a certain instant of explosive processes. [1,2] In order to penetrate dense materials, a high-energy intense-current electron beam is focused onto a high-Z target to produce x-ray pulse from bremmstrahlung radiation. [3,4] For the purpose of resolving fine details in the object, narrow temporal width and small spot size are required for the x-ray source, which help to reduce the motion blur and the geometry blur, respectively. The pulse width is typically tens of nanoseconds and usually cannot be further shortened due to limitation of dose needed. The spot size of the x-ray source is often quoted as the evaluation of the resolving ability for a particular flash-radiography machine.

During the passing decades, various techniques have been developed for the x-ray spot size measurement. [5-8] Different definitions of spot size were introduced simultaneously, such as the full-width at half maximum (FWHM) [5,6], the Los Alamos National Laboratory (LANL) spot size [7], the Atomic Weapon Establishment (AWE) spot size [8] as well as the limiting resolution [9]. In this paper, the method of pinhole imaging is applied to measure the x-ray spot size of the Dragon-I LIA [10,11]. This method possesses an advantage of obtaining the whole two-dimensional spatial distribution of the radiographic spot, which generally exhibits azimuthal asymmetry. Both the FWHM of the x-ray spot and the LANL spot size based on modulation transfer function (MTF) are calculated.

## 2. Principle

The point-spread function (PSF) of an ideal pinhole is a delta-function. Consequently, the result obtained by the pinhole measurement is exact the PSF of source spot after dividing its spatial dimensions by the magnification of imaging setup. When the PSF is projected along a direction, the obtained result is called the line-spread function (LSF). Here, we take the projection of y-direction for example, the LSF, L(x), can be expressed as

$$L(x) = \int_{-\infty}^{+\infty} p(x,y)\mathrm{d}y, \quad (1)$$

where p(x,y) is the PSF which describes a normalized two-dimensional spatial distribution.

A simple way to characterize spot size is to use FWHM. It should be noticed that FWHM of PSF is generally unequal to that of the corresponding LSF except for the Gaussian distribution. Table 1 compares FWHMs of the PSF and the LSF for typical distributions which are common referred to, including uniform disk (KV), Gaussian (GS), Bennett (BNT) and Quasi-Bennett (QBNT) [9]. They are listed in a trend from a more centralized distribution to more expanding one. It can be seen that the radio of FWHM(LSF) to FWHM(PSF) increases as the peak of the function becomes sharper while the wing becomes broader.

Table 1. FWHMs of PSF and LSF for different distributions

| Distribution | PSF(p(x,y)) | FWHM(PSF) | LSF(L(x)) | FWHM(LSF) | FWHM(LSF)/FWHM(PSF) |
|---|---|---|---|---|---|
| KV | $\begin{cases} \dfrac{1}{\pi R_0^2}, & x^2+y^2 \le R_0^2 \\ 0, & x^2+y^2 > R_0^2 \end{cases}$ | $2R_0$ | $\begin{cases} \dfrac{2\sqrt{a^2-x^2}}{\pi R_0^2}, & x^2 \le R_0^2 \\ 0, & x^2 > R_0^2 \end{cases}$ | $\sqrt{3}R_0$ | 0.866 |
| GS | $\dfrac{1}{\pi a^2}\exp\left(-\dfrac{x^2+y^2}{a^2}\right)$ | $2\sqrt{\ln 2}\,a$ | $\dfrac{1}{\sqrt{\pi}a}\exp\left(-\dfrac{x^2}{a^2}\right)$ | $2\sqrt{\ln 2}\,a$ | 1 |
| BNT | $\dfrac{1}{\pi a^2}\left(1+\dfrac{x^2+y^2}{a^2}\right)^{-2}$ | $2\sqrt{\sqrt{2}-1}\,a$ | $\dfrac{1}{2}\left(1+\dfrac{x^2}{a^2}\right)^{-\frac{3}{2}}$ | $2\sqrt{2^{2/3}-1}\,a$ | 1.191 |
| QBNT | $\dfrac{1}{2\pi a^2}\left(1+\dfrac{x^2+y^2}{a^2}\right)^{-\frac{3}{2}}$ | $2\sqrt{2^{2/3}-1}\,a$ | $\dfrac{a}{\pi(x^2+a^2)}$ | $2a$ | 1.305 |

Apparently, the characterization by spot size by FWHM does not take into account the spatial distribution. The MTF is used to analyze each imaging component as a low pass filter for spatial information. In LANL definition, the size of an actual spot ($D_{\mathrm{LANL}}$) is equal to the diameter of an equivalent uniform disk which has the same spatial frequency at half of the MTF peak value. The relation is given as [7]

$$D_{\mathrm{LANL}} = \dfrac{0.705}{f_{50\%\mathrm{MTF}} \cdot M}, \quad (2)$$

where $f_{50\%\mathrm{MTF}}$, with the unit of inverse dimension, is the spatial frequency of the image LSF, from which the blur of image recording system should be deducted; M is the geometrical magnification of the imaging setup.

The MTF is a two-dimensional surface in spatial frequency space, which is defined as the modulus of Fourier transform of the PSF:

$$F(f_x, f_y) = \left| \int_{-\infty}^{+\infty} dx \int_{-\infty}^{+\infty} p(x,y) \exp\left[-i2\pi(f_x x + f_y y)\right] dy \right|, \quad (3)$$

where $F(f_x, f_y)$ is the MTF, $f_x$ and $f_y$ represent spatial frequencies. The slice $f_y = 0$ of the MTF can be expressed as

$$F(f_x, 0) = \left| \int_{-\infty}^{+\infty} dx \int_{-\infty}^{+\infty} p(x,y) \exp(-i2\pi f_x x) dy \right|$$
$$= \left| \int_{-\infty}^{+\infty} L(x) \exp(-i2\pi f_x x) dx \right|, \quad (4)$$

which is actually the modulus of Fourier transform of LSF corresponding to the y-direction projection of PSF.

## 3. Experimental setup

The experimental setup for x-ray spot size measurement based on the pinhole imaging technique is shown in Fig. 1. A tungsten cube object with a pinhole perforated is laid between the conversion target and the image recording system. The placement of pinhole should be carefully adjusted to make sure that the pinhole and the x-ray path have identical axes. The diameter of the pinhole is $d = 0.47$ mm, and the thickness along the axis is $L = 65$ mm. The distance between the x-ray source plane and the rear facet of the pinhole is $a = 1178$ mm while that between the rear facet of the pinhole and the CsI scintillator screen is $b = 5087$ mm. The magnification is therefore $M = b/a = 4.318$. The energy of electron beam is kept in the range of $E = 18.9 \sim 19.2$ MeV, and the current is $I = 2.4 \sim 2.5$ kA.

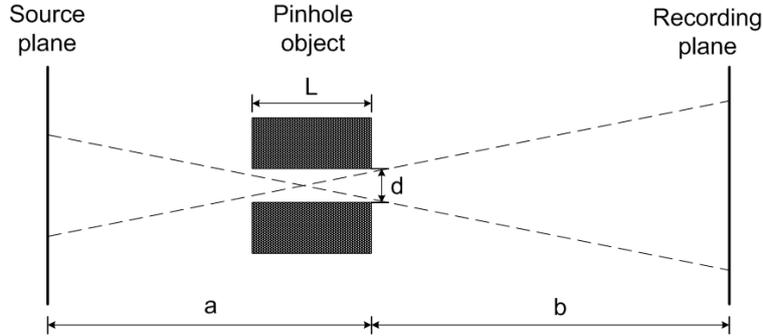

Fig. 1 Experimental step of the pinhole technique

After being accelerated, the electron beam is transported through the downstream transport section of the LIA and finally focused onto the x-ray conversion target. In the downstream transport section, the focusing solenoid (FS), together with two other short solenoids (SS1 and SS2), is tuned by changing the loading current. Efforts are made to maintain a constant state of the electron beam. During the experiment, the energy of electron beam is kept in the range of $E = 18.9 \sim 19.2$ MeV, and the current in the range of $I = 2.4 \sim 2.5$ kA.

## 4. Results and discussions

Experiments are performed to measure the x-ray spot size based on the pinhole imaging technique. The experimental results of FWHM are given in Table 2, where FWHMs of the PSF and two LSFs (including x-direction and y-direction projections) are given for each measurement. Typical images of the x-ray source spot are shown in Fig. 2. Both the boundaries of 50% peak value (in back curve) and 10% peak value (in white curve) are lined out. The FWHM of the spot image is given by the diameter of an equivalent circle that has the same area as the boundary of 50% peak value. Then the image FWHM is divided by the magnification to obtain the FWHM of source PSF (p(x,y)). After that, two LSFs are calculated by projecting the spot image in two perpendicular directions. As shown in Fig. 3, the LSF along the x-direction is compared with typical distributions in theory. It is seen that the FWHM of PSF decreases as tuning the currents of FS, SS1 and SS2, meantime, the wing of LSF becomes more expanding. The ratio value of FWHM(L(x))/FWHM(p(x,y)) is calculated to be 1.18 for No.#2, which is close to the value of BNT, and 1.33 for No.#6, which is close to the value of QBNT.

Table 2. Experimental FWHMs of the PSF and the LSF

| No. | E /MeV | I /kA | FS /A | SS1 /A | SS2 /A | FWHM(p(x,y)) /mm | FWHM(L(x)) /mm | FWHM(L(y)) /mm |
|---|---|---|---|---|---|---|---|---|
| #1 | 19.0 | 2.5 | 570 | 320 | 190 | 1.48 | 1.96 | 1.68 |
| #2 | 18.9 | 2.5 | 570 | 320 | 230 | 1.62 | 1.91 | 1.99 |
| #3 | 18.9 | 2.5 | 590 | 320 | 190 | 1.45 | 1.90 | 1.71 |
| #4 | 19.2 | 2.4 | 600 | 280 | 190 | 1.38 | 1.77 | 1.72 |
| #5 | 19.0 | 2.5 | 600 | 280 | 190 | 1.32 | 1.83 | 1.75 |
| #6 | 19.0 | 2.4 | 600 | 280 | 190 | 1.33 | 1.77 | 1.72 |

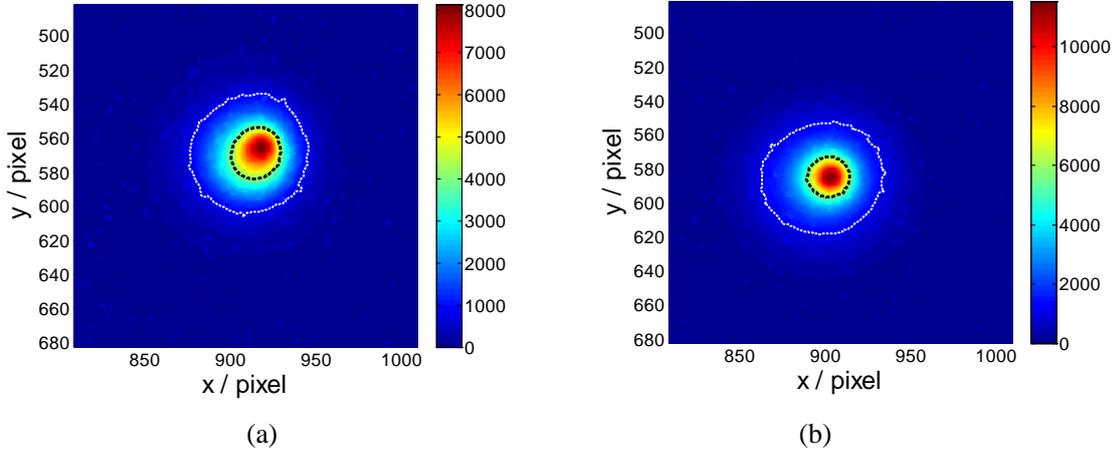

(a)     (b)

Fig. 2 Experimental images of the x-ray source spot. (a) No. #2 with the PSF FWHM of 1.62 mm; (b) No. #6 with the PSF FWHM of 1.33 mm. The black curve is the boundary of 50% peak value and the white curve is that of 10% peak value.

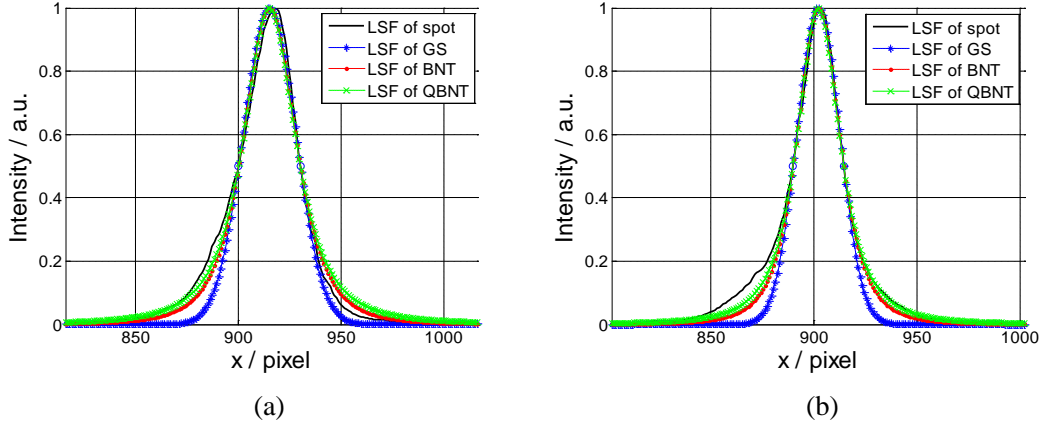

Fig. 3 Comparison of measured LSFs along x-direction with theoretical functions. (a) No. #2 with the LSF FWHM of 1.91 mm; (b) No. #6 with the LSF FWHM of 1.77 mm.

The LANL spot size is also calculated by making Fourier transform of the image LSFs and finding the spatial frequency at half of the MTF maximum. In order to make a correction of the blur of the image recording system, a 10-mm thick tungsten plate is placed in contact with the scintillator and then the penumbral image of the edge is recorded, which actually denotes the edge spread function (ESF). The MTF of blur is finally obtained after differentiating the ESF and further making Fourier transform of its LSF.

According to Eq. (2), $D_{LANL}$ is calculated and listed in Table 3. For each measurement, the results of $D_{x,LANL}$ and $D_{y,LANL}$ correspond to the y-direction projection and the x-direction projection, respectively. Fig. 4 shows the MTF curves of the two measurements mentioned above, including MTFs of the image LSF, the blur and the image LSF with blur deducted. It seems that $D_{LANL}$ is not directly related to the spot FWHM. For No. #2, the LANL spot sizes ($D_{x,LANL} = 3.35$ mm, $D_{y,LANL} = 3.36$ mm) are both smaller than the results of No. #6 ($D_{x,LANL} = 3.78$ mm, $D_{y,LANL} = 3.39$ mm), while their spot FWHMs show a completely reverse relation. The reason is due to the fact that $D_{LANL}$ defined from spatial frequency and MTF has an intrinsic and close relation with the spatial distribution of source spot.

Table 3. Experimental results of the LANL spot size

| No. | $f_{x,50\% MTF}$ / mm$^{-1}$ | | $D_{x,LANL}$ / mm | $f_{y,50\% MTF}$ / mm$^{-1}$ | | $D_{y,LANL}$ / mm |
|---|---|---|---|---|---|---|
| | with blur | without blur | | with blur | without blur | |
| #1 | 0.0370 | 0.0440 | 3.71 | 0.0387 | 0.0479 | 3.41 |
| #2 | 0.0399 | 0.0488 | 3.35 | 0.0400 | 0.0486 | 3.36 |
| #3 | 0.0375 | 0.0449 | 3.64 | 0.0410 | 0.0496 | 3.29 |
| #4 | 0.0377 | 0.0457 | 3.58 | 0.0403 | 0.0488 | 3.34 |
| #5 | 0.0366 | 0.0432 | 3.78 | 0.0401 | 0.0484 | 3.37 |
| #6 | 0.0364 | 0.0432 | 3.78 | 0.0399 | 0.0481 | 3.39 |

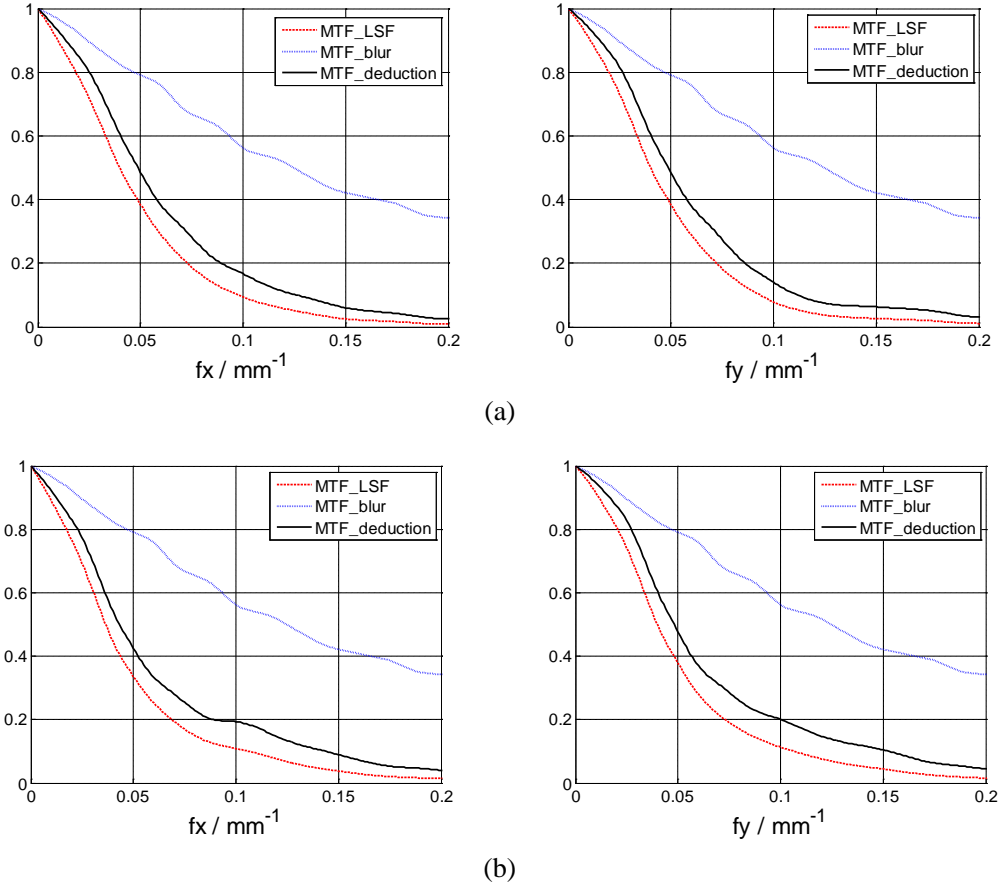

Fig. 4 MTF curves of experimental spot images. (a) No. #2; (b) No. #6.

## 5. Conclusion

The pinhole imaging technique is applied to measure the x-ray source spot size of Dragon-I LIA, by which a two-dimensional spatial distribution of the spot image is obtained. Both the FWHM of the spot and the size in LANL definition are calculated. The PSF FWHM is given by the diameter of a disk which has an area equal to that of the boundary at 50% PSF peak value. The LANL spot size is calculated through projecting the image PSF to get the LSF and further making Fourier transform to obtain the MTF. Transportation and focusing of the electron beam are tuned through adjusting currents of the solenoids in the downstream section. Experimental results indicate that compared with FWHM the LANL spot size shows to be more sensitive to the variation of source spatial distribution.